\documentclass[preprint,showpacs,preprintnumbers,amsmath,amssymb]{revtex4}

\usepackage{dcolumn, bm, latexsym,amsmath,amsfonts,amssymb,graphicx, textcomp, multirow}

\begin{document}

\preprint{}

\title{Spin-orbit induced interference in polygon-structures}

\author{Marc J. van Veenhuizen}
\email{mvanveen@mit.edu}
\altaffiliation[Present address:]{Department of Physics, MIT}
\affiliation{NTT Basic Research Laboratories, 3-1 Morinosato-Wakamiya, Atsugi-shi, Kanagawa-ken, 243-0198, Japan}
\author{Takaaki Koga}
\altaffiliation[Present address:]{Graduate School of Information Science and Technology, Hokkaido University, N-14, W-9, Sapporo, Hokkaido, 060-0814, Japan}
\affiliation{PRESTO, Japan Science and Technology Agency (JST)}
\affiliation{NTT Basic Research Laboratories, 3-1 Morinosato-Wakamiya, Atsugi-shi, Kanagawa-ken, 243-0198, Japan}
\author{Junsaku Nitta}
\affiliation{NTT Basic Research Laboratories, 3-1 Morinosato-Wakamiya, Atsugi-shi, Kanagawa-ken, 243-0198, Japan}
\affiliation{CREST, Japan Science and Technology Agency (JST)}

\date{\today}

\begin{abstract}
We investigate the spin-orbit induced spin-interference pattern of ballistic electrons travelling along \textit{any} regular polygon. It is found that the spin-interference depends strongly on the Rashba and Dresselhaus spin-orbit constants as well as on the sidelength and alignment of the polygon. We derive the analytical formulae for the limiting cases of either zero Dresselhaus or zero Rashba spin-orbit coupling, including the result obtained for a circle. We calculate the nonzero Dresselhaus and Rashba case numerically for the square, triangle, hexagon, and circle and discuss the observability of the spin-interference which can potentially be used to measure the Rashba and Dresselhaus coefficients.
\end{abstract}


\maketitle

\section{\label{sec:intro}Introduction}

In recent years, much research has been devoted to study means to manipulate the spin of electrons. A prime candidate for spin-manipulation is the Rashba spin-orbit interaction \cite{Rashba60, BychkovRashba84} which arises in certain semiconductor quantum wells such as InGaAs heterostructures as a result of structural inversion asymmetry. The strength of the spin-orbit interaction depends on the design of the heterostructures \cite{Koga02} and can be dynamically altered by applying a gate voltage \cite{Nitta97}, hence offering a way to manipulate the spin in a controlled fashion. \newline
Another spin-orbit interaction present in for instance GaAs is the Dresselhaus spin-orbit interaction \cite{Dresselhaus1955} which stems from the absence of bulk inversion symmetry. Recently \cite{Kato2003} it has been shown for bulk GaAs structures that induced uniform strain affects the Dresselhaus spin-splitting hence giving another way to engineer spin manipulation.\newline
In this article, which is a justification and generalization of our recently introduced proposal \cite{Koga2004}, we will further develop our idea to study spin manipulation by means of the Rashba and Dresselhaus interaction via spin-interference effects on the conductance. We will treat the Rashba and Dresselhaus spin-splitting on an equal footing and we will derive the interference pattern resulting from their combined effect on the spin of the electron.\newline
For ring-structures the interference effect due to spin-orbit interaction has been well investigated \cite{Meir89, Mathur92, Aronov93, Qian94, Nitta99} and for square loops it has recently be shown \cite{Berc2004} that interference due to Rashba spin-orbit coupling can lead to electron localization.  We will discuss the interference pattern due to the Rashba \textit{and} Dresselhaus spin-orbit interaction of electrons travelling along \textit{polygons}. In this way we are able to propose a scheme how to experimentally obtain the Rashba and Dresselhaus parameters from the spin-interference.
The electrons will be treated as ballistic wave packets with a spin travelling both clockwise and counter-clockwise along the sides of a regular polygon which is positioned in the $\hat{\bm{x}}$-$\hat{\bm{y}}$-plane, the 2DEG of a semiconductor heterostructure. Since the spin-orbit interaction depends on the travel direction, the spins of the electrons will rotate along different directions for each side of the polygon with a precession angle that depends both on the sidelength as well as on the strength of the Rashba and Dresselhaus spin-splitting. For certain precession angles the counter-clockwise and clockwise spin waves interfere completely destructive or constructive after a full rotation, which corresponds respectively to a large and small conductivity. We include an Aharonov-Bohm (AB) phase difference between the counter-clockwise and clockwise spin waves and derive that for an unpolarized beam the spin-interference affects the amplitude of the AB oscillation. The procedure we use to calculate the spin-interference transforms the spin by a series of quantum-mechanical rotations corresponding to the various sides of the polygon.

\section{\label{sec:quantum1}Interference intensity}

\begin{figure}[htb]
\begin{center}
\resizebox{80mm}{!}{\includegraphics{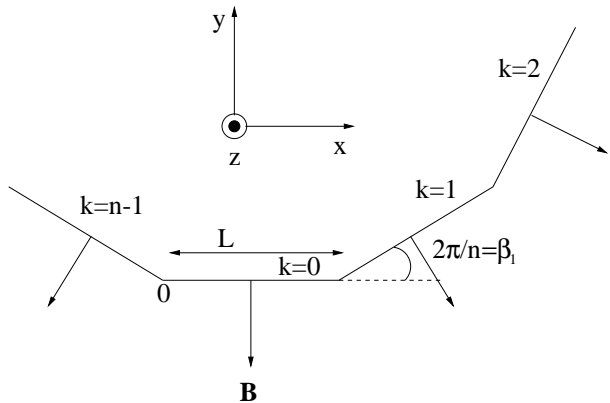}}
\caption{\label{fig:polygon}Polygon with $n$ vertices. The effective magnetic field direction changes as a function of the side $k$.}
\end{center}
\end{figure}

Consider the $n$-sided regular polygon of Fig.\ \ref{fig:polygon} which the electron traverses both clockwise and counter-clockwise simultaneously.
Let us concentrate first on the counter-clockwise travelling part. The spin-orbit interaction manifests itself as an effective magnetic field about which the spin precesses. This effective magnetic field is constant along a side but differs for different sides. We assume for the moment that the electron will start at the beginning of the $k=0$ side indicated by the $0$ in Fig.\ \ref{fig:polygon} which we will call the \textit{origin}. We will denote the quantum mechanical rotation of the spin along a side $k$ with $R_k$ and the precession angle (of which $R_k$ is a function) with $s$. Suppose that the initial spin polarisation direction is along the $z$-axis. We thus have a $\bm{J}^2$ and $J_z$ eigenstate $\left|{\frac{1}{2},m}\right\rangle\equiv \left|{m}\right\rangle$. The wavefunction after travelling along the side $k$ then reads
\begin{equation}
R_k\left|{m}\right\rangle = \sum_{m'}D_{m'm}^{1/2}(R_k)\left|{m'}\right\rangle,
\end{equation}
where $D_{m'm}^{1/2}(R_k)$ is the rotation matrix \cite{Rose57} corresponding to the rotation operator $R_k$. Using the group multiplication law for the rotation matrices repeatedly, we find for the spin wave function after completing the polygon counter-clockwise
\begin{equation}\label{eq:wavefcnccwrotated}
A\left|{m}\right\rangle = \sum_{m'}D_{m'm}^{1/2}(A)\left|{m'}\right\rangle,
\end{equation}
where the rotation $A$ is given by
\begin{equation}\label{eq:rotationA}
\begin{aligned}
A &= R_{n-1}\cdots R_0 \\
&= e^{-n\zeta\hat{\bm{\xi}}\cdot\bm{J}/\hbar}.
\end{aligned}
\end{equation}
The rotation $A$ is characterised by the rotation direction unit vector $\hat{\bm{\xi}}$ and angle $n\zeta$ which we will calculate analytically later for some limiting cases.
The clockwise going spin wave function reads
\begin{equation}
\begin{aligned}
A^{-1}\left|{m}\right\rangle &= \sum_{m'}D_{m'm}^{1/2}(A^{-1})\left|{m'}\right\rangle
&= \sum_{m'}D_{mm'}^{1/2\ast}(A)\left|{m'}\right\rangle,
\end{aligned}
\end{equation}
where the last step follows from the unitarity of the Wigner $D$-functions and $\ast$ denotes the complex conjugate.
The total final state becomes
\begin{equation}\label{eq:finalwave}
\left|{\psi}\right\rangle = \frac{1}{\sqrt{2}}A\left|{m}\right\rangle + \frac{1}{\sqrt{2}}A^{-1}\left|{m}\right\rangle,
\end{equation}
yielding for the interference intensity in the origin
\begin{equation}\label{eq:probability1}
|\psi(0)|^2 = 1+ Re\left(D_{mm}^{1/2}(A^2)\right).
\end{equation}
We now generalize by starting with a spin polarised in an arbitrary direction, which we however can decompose into a linear combination of spin-up and spin-down along the original z-axis
\begin{equation}\label{eq:arbitraryspin}
\left|{\nearrow m''}\right\rangle \equiv a\left|{\uparrow}\right\rangle + b\left|{\downarrow}\right\rangle.
\end{equation}
Furthermore, we include the effect of a magnetic field causing an Aharonov-Bohm phase difference $\delta$ between the counter-clockwise and clockwise travelling wave (we ignore the effect of the magnetic field on the spin). We find for the total final state
\begin{equation}\label{eq:finalwavephasearbitraryspin}
\begin{aligned}
\left|{\psi}\right\rangle &= e^{i\delta}\frac{1}{\sqrt{2}}A\left|{\nearrow m''}\right\rangle + \frac{1}{\sqrt{2}}A^{-1}\left|{\nearrow m''}\right\rangle,
\end{aligned}
\end{equation}
and for the interference intensity
\begin{equation}
\begin{aligned}
|\psi(0)|^2 &= |a|^2\left(1+Re\left(e^{i\delta}D_{\frac{1}{2}\frac{1}{2}}^{\frac{1}{2}}(A^2)\right)\right)\\
&\quad +|b|^2\left(1+Re\left(e^{i\delta}D_{-\frac{1}{2}-\frac{1}{2}}^{\frac{1}{2}}(A^2)\right)\right)\\
&\quad +2Re\left(a^*b\left(e^{-i\delta}D_{-\frac{1}{2}\frac{1}{2}}^{\frac{1}{2}*}(A^2)+e^{i\delta}D_{\frac{1}{2}-\frac{1}{2}}^{\frac{1}{2}*}(A^2)\right)\right).
\end{aligned}
\end{equation}
In the following we will frequently make use of Euler parameters to express rotations, for a discussion of Euler parameters see, for instance, \cite{Gold80}. 
The interference intensity in terms of Euler parameters reads
\begin{equation}
\begin{aligned}
|\psi(0)|^2 &= 1 + e_0\cos\delta + (|a|^2-|b|^2)e_3\sin\delta\\
& \quad + 4\left[Re(a^{*}b)e_1 + Im(a^{*}b)e_2\right]\sin\delta.
\end{aligned}
\end{equation}
If we average over the spin polarisation the intensity reduces to
\begin{equation}\label{eq:averageprobability}
\overline{|\psi(0)|^2} = 1 + e_0(A^2)\cos\delta.
\end{equation}
In the following Section we will discuss the various contributions to the rotation $A^2$ and seek a closed expression for it.
\section{rotation}
The important spin-orbit contributions to the Hamiltonian are the Rashba and linear Dresselhaus terms, for a $[001]$ 2DEG given by, respectively,
\begin{equation}\label{eq:sohamiltonian}
\begin{aligned}
H_R &= \alpha_R(\hat{\bm{x}}k_y - \hat{\bm{y}}k_x)\\
H_{D1} &= \alpha_{D1}(-\hat{\bm{x}}k_x+\hat{\bm{y}}k_y)\\
\end{aligned}
\end{equation}
Their effect on the spin can be described as a rotation about the sum of the effective magnetic fields of the individual terms, the (dimensionless) direction of which is given by the vector
\begin{equation}\label{eq:totalB}
\bm{B} = s_R\left(\begin{array}{c}\sin\beta\\-\cos\beta\\0\end{array}\right)
+s_{D1}\left(\begin{array}{c}-\cos\beta\\\sin\beta\\0\end{array}\right)
\end{equation}
which follows from Eq.\ (\ref{eq:sohamiltonian}) by expressing the wavevector in the angle $\beta$ with the $x$-axis. The parameter $s_i,\;i=R,D1$ is the angle through which the spin would rotate about the effective magnetic field direction $B_i$ in the absence of the other. Hence $s_i$ determines the relative weight of that particular term in the sum. The precession angle about the total effective magnetic field is the modulus of the vector Eq.\ (\ref{eq:totalB})
\begin{equation}
s(\beta) = |\bm{B}|.
\end{equation}
The Euler parameters then read
\begin{equation}
\begin{aligned}
e_0 &= \cos\frac{s(\beta)}{2}\\
\bm{e} &= \hat{\bm{B}}\sin\frac{s(\beta)}{2}.
\end{aligned}
\end{equation}
For a polygon, the angle $\beta$ assumes only the discrete values $\beta_k=2\pi k/n$ where $n$ is the number of vertices of the polygon and $k$ ranges from $0$ to $n-1$. Notice that we restrict for the moment 1 edge of the polygon to be positioned along the $x$-axis (as in Fig.\ \ref{fig:polygon}), a condition on which we will comment extensively later on.
In the expectation that rotations along successive polygon edges will 
partially cancel each other we decompose the rotation along a side into 
Euler angles $(\varphi,\theta,\psi)$: $R_k = 
R_z(\varphi)R_y(\theta)R_z(\psi)$ (We use the socalled $zyz$-convention).
We find for the Euler angles
\begin{equation}
\begin{aligned}
\theta &= sign(B)s(\beta_k)\\
\varphi &= -\psi\\
\tan\psi &= \frac{s_R\sin\beta_k-s_{D1}\cos\beta_k}{-s_R\cos\beta_k+s_{D1}\sin\beta_k},
\end{aligned}
\end{equation}
where we have included the overall sign of $\bm{B}$ in the Euler angle $\theta$ since it is divided out in the expression for the other Euler angles. The Euler angles become particularly simple for $s_{D1}=\pm s_R$ for then $\psi=-\varphi=\pi/4,3\pi/4$, $\theta=\sqrt{2}s_R(\sin\beta_k\mp \cos\beta_k)$, respectively, and the total rotation Eq.\ (\ref{eq:rotationA}) becomes
\begin{equation}\label{eq:aRisaD1}
A = R_z\left(-\frac{(3)\pi}{4}\right)R_y\left(\sqrt{2}s_R\sum_{k=0}^{n-1}\sin\beta_k\mp\cos\beta_k\right)R_z\left(\frac{(3)\pi}{4}\right) = 1.
\end{equation}
Hence we always have a maximum for $s_{D1}=\pm s_R$.\newline
\subsection{$\alpha_{D1}=0$ or $\alpha_R=0$}
Consider now the limit $\alpha_{D1}=0$. We find for the Euler angles
\begin{equation}
(\varphi,\theta,\psi) = (\beta_k,s,-\beta_k),
\end{equation}
where we defined a new variable
\begin{equation}
\beta_k \equiv \frac{2\pi k}{n},
\end{equation}
and where we have written $s$ for $s_R$.
The rotation $R_k$ we may then write as
\begin{equation}
R_k = R_z(\beta_k)R_y(s)R_z(-\beta_k).
\end{equation}
The product of two adjacent rotations in Eq.\ (\ref{eq:rotationA}) takes the form
\begin{equation}
R_{k+1}R_k = R_z(\beta_{k+1})R_y(s)R_z(-\beta_1)R_y(s)R_z(-\beta_k).
\end{equation}
The total rotation Eq.\ (\ref{eq:rotationA}) then simplifies to
\begin{equation}\label{eq:derivedtotalrotation}
\begin{aligned}
A = R_{n-1}\cdots R_0 
&= R_z(\beta_n)\left[R_z(-\beta_1)R_y(s)\right]^n \\
&= -R_{\xi}(n\zeta),
\end{aligned}
\end{equation}
where we introduced a new rotation
\begin{equation}\label{eq:Rzeta}
R_{\xi}(\zeta) \equiv R_z(-\beta_1)R_y(s).
\end{equation}
The interpretation of the new rotation Eq. (\ref{eq:Rzeta}) is straightforward: first a rotation about $y$ and then, since the magnetic field rotates counter-clockwise, the spin vector rotates clockwise about $z$. In terms of Eulerparameters the rotation Eq. (\ref{eq:Rzeta}) reads
\begin{subequations}
\begin{eqnarray}
\label{eq:relationzetas}
&e_0& = \cos\frac{\beta_1}{2}\cos\frac{s}{2} = \cos\frac{\zeta}{2} \\
&e_1& = \sin\frac{\beta_1}{2}\sin\frac{s}{2} = \hat{\xi}_x\sin\frac{\zeta}{2} \\
&e_2& = \cos\frac{\beta_1}{2}\sin\frac{s}{2} = \hat{\xi}_y\sin\frac{\zeta}{2} \\
&e_3& = -\sin\frac{\beta_1}{2}\cos\frac{s}{2} = \hat{\xi}_z\sin\frac{\zeta}{2},
\end{eqnarray}
\end{subequations}
yielding for the rotation-direction
\begin{equation}
\hat{\bm{\xi}} = \frac{1}{\sqrt{1-\cos^2\frac{\beta_1}{2}\cos^2\frac{s}{2}}}\left(\begin{array}{r}
\sin\frac{\beta_1}{2}\sin\frac{s}{2} \\
\cos\frac{\beta_1}{2}\sin\frac{s}{2} \\
-\sin\frac{\beta_1}{2}\cos\frac{s}{2}
\end{array}\right).
\end{equation}
The Euler parameters of the square of the total rotation Eq. (\ref{eq:rotationA}) (i.e. $A^2$)  are
\begin{equation}\label{eq:derivedtotalrotationEpar}
\begin{aligned}
&e_0 = \cos n\zeta, \\
&\bm{e} = \hat{\bm{\xi}}\sin n\zeta.
\end{aligned}
\end{equation}
From Eq. (\ref{eq:averageprobability}) we can directly read off the condition for a maximum respectively a minimum in the interference intensity in terms of the Euler parameters
\begin{subequations}
\begin{eqnarray}
\label{eq:rootrelation1}
&e_0 = 1 \\
\label{eq:rootrelation2}
&e_0 = -1.
\end{eqnarray}
\end{subequations}
Eqs. (\ref{eq:rootrelation1}) respectively (\ref{eq:rootrelation2}) together with Eq. (\ref{eq:derivedtotalrotationEpar}) specify the angles $\zeta$ for which we have a maximum respectively a minimum. Combining Eq. (\ref{eq:rootrelation1}) with Eq. (\ref{eq:relationzetas}) we find a maximum intensity if
\begin{equation}\label{eq:rootcondition1}
\cos s = \frac{\cos\frac{2m\pi}{n}+1}{\cos^2\frac{\pi}{n}}-1,\;\; m\in\left[1,\frac{n}{2}\right].
\end{equation}
From Eq. (\ref{eq:rootrelation2}) together with Eq. (\ref{eq:relationzetas}) we find a minimum intensity if
\begin{equation}\label{eq:rootcondition2}
\cos s = \frac{\cos\frac{(2m+1)\pi}{n}+1}{\cos^2\frac{\pi}{n}}-1,\;\; m\in\left[1,\frac{n-1}{2}\right].
\end{equation}
The restrictions for $m$ follow from $\cos s\in[-1,1]$. \newline
From Eq. (\ref{eq:relationzetas}) we see that if (and only if) $s\rightarrow  s +2\pi$, then $\zeta \rightarrow \zeta +2\pi$. The rotation $A^2$ is characterised by $n2\zeta$ which transforms then as $n2\zeta \rightarrow n2\zeta + n4\pi = n2\zeta$. Thus $A^2$ is periodic in $s$ with period $2\pi$, ergo is the interference intensity. That is the reason why we could rewrite the conditions to get extrema for $\cos s/2$ into conditions for $\cos s$ without loosing any information.\newline
Apart from the round trip interference which happens at the entrance lead of the polygon we could also imagine a lead halfway of the polygon causing interference in the forward direction. It can be readily observed that the forward interference intensity is given by the same expressions as the backward interference, except for the substitution $A^2\rightarrow A$. The intensity halfway attains maxima and minima, respectively, for
\begin{subequations}
\begin{eqnarray}
\cos\frac{(2m+1)\pi}{n} &=& \cos\frac{\pi}{n}\cos\frac{s}{2},\;\; m\in\left[1,\frac{n}{4}\right] \\
\cos\frac{2m\pi}{n} &=& \cos\frac{\pi}{n}\cos\frac{s}{2},\;\; m\in\left[1,\frac{n}{4}\right].
\end{eqnarray}
\end{subequations}
Notice that for $n$ odd we cannot rewrite the condition for $\cos s/2$ in a condition for $\cos s$ since the period in $s$ is $4\pi$ for $n$ odd.\newline
If the starting point of the spin wave function is not the origin but, say, $p$ radians removed from the origin $0$ (on the $k=0$ side), the total rotation becomes
\begin{equation}
A' = R_y(p)AR_y^{-1}(p),
\end{equation}
which is merely a similarity transformation and transforms the spin to a different basis. Since we average over the spin a similarity transformation does not change the result for the intensity. If the polygon has not one side aligned along the $\hat{\bm{x}}$-direction but tilted at some angle $\chi$ then we have the substitution $\beta_k\rightarrow\beta_k+\chi$. Again, the total rotation will be changed by a similarity transformation
\begin{equation}
A' = R_z(\chi)AR_z^{-1}(\chi),
\end{equation}
unaltering the results for the intensity. Things \textit{change} however if both the Rashba and Dresselhaus interaction play a role, as shown in Section \ref{subsec:aRaD1neq0} since then adjacent rotations do not cancel each other neatly.\newline
The Rashba $B$-field becomes the linear Dresselhaus $B$-field if we make the substitution $\beta_k\rightarrow\beta_k-\pi/2$. The total rotation Eq. (\ref{eq:rotationA}) once again changes by a similarity transformation
\begin{equation}
A' = R_z(-\pi/2)AR_z^{-1}(-\pi/2).
\end{equation}
Hence the derived results hold as well in the limit $\alpha_{R}=0, \alpha_{D1}\neq0$.
\newline
We will now rewrite Eq. (\ref{eq:averageprobability}) in terms of physical variables. According to \cite{DattaDas90} the spin precession angle $s$ in a one dimensional ballistic channel with Rashba-SO coupling is given by
\begin{equation}\label{eq:angleRashbaLength}
s = \frac{2\alpha m^{*}}{\hbar^2}\times L,
\end{equation}
with $\alpha$ the Rashba coefficient, $m^*$ the effective electron mass and $L$ the sidelength of the polygon. Using Eq. (\ref{eq:relationzetas}) to express $\zeta$ in $s$ we find for the averaged interference intensity Eq. (\ref{eq:averageprobability})
\begin{equation}\label{eq:probphyspolygon}
\begin{aligned}
&\overline{|\psi(0)|^2} = \\
&1 + \cos \left(2n\arccos\left[\cos\frac{\pi}{n}\cos\left(\frac{2\alpha m^*}{\hbar^2}\frac{1}{2}L\right)\right]\right)\cos\delta,
\end{aligned}
\end{equation}
where we substituted Eq. (\ref{eq:angleRashbaLength}) for the precession angle $s$. An interesting limit is to let the number of vertices $n$ go to infinity and at the same rate reducing the sidelength $L$, thereby obtaining a circle. From Fig.\ \ref{fig:polygon} we derive the relation
\begin{equation}\label{eq:circumradiusnL}
\sin\frac{\pi}{n} = \frac{L}{2R},
\end{equation}
with $R$ the circumradius. We can express $n$ in terms of $L$ by using Eq. (\ref{eq:circumradiusnL}), and taking the limit $L\rightarrow 0$ we find for the intensity
\begin{equation}\label{eq:probphyscircle}
\begin{aligned}
&\overline{|\psi(0)|^2}\mid_{\rm{circle}} = \\
&1 + \cos \left(2\pi\sqrt{1+\left(\frac{2\alpha m^*}{\hbar^2}R\right)^2}\right)\cos\delta.
\end{aligned}
\end{equation}
The intensity halfway is obtained from Eqs. (\ref{eq:probphyspolygon}, \ref{eq:probphyscircle}) by dividing the argument of the cosine by $2$.
\subsection{$\alpha_R,\;\alpha_{D1} \neq 0$\label{subsec:aRaD1neq0}}
\begin{figure}[h]
\begin{center}
\resizebox{100mm}{!}{\includegraphics{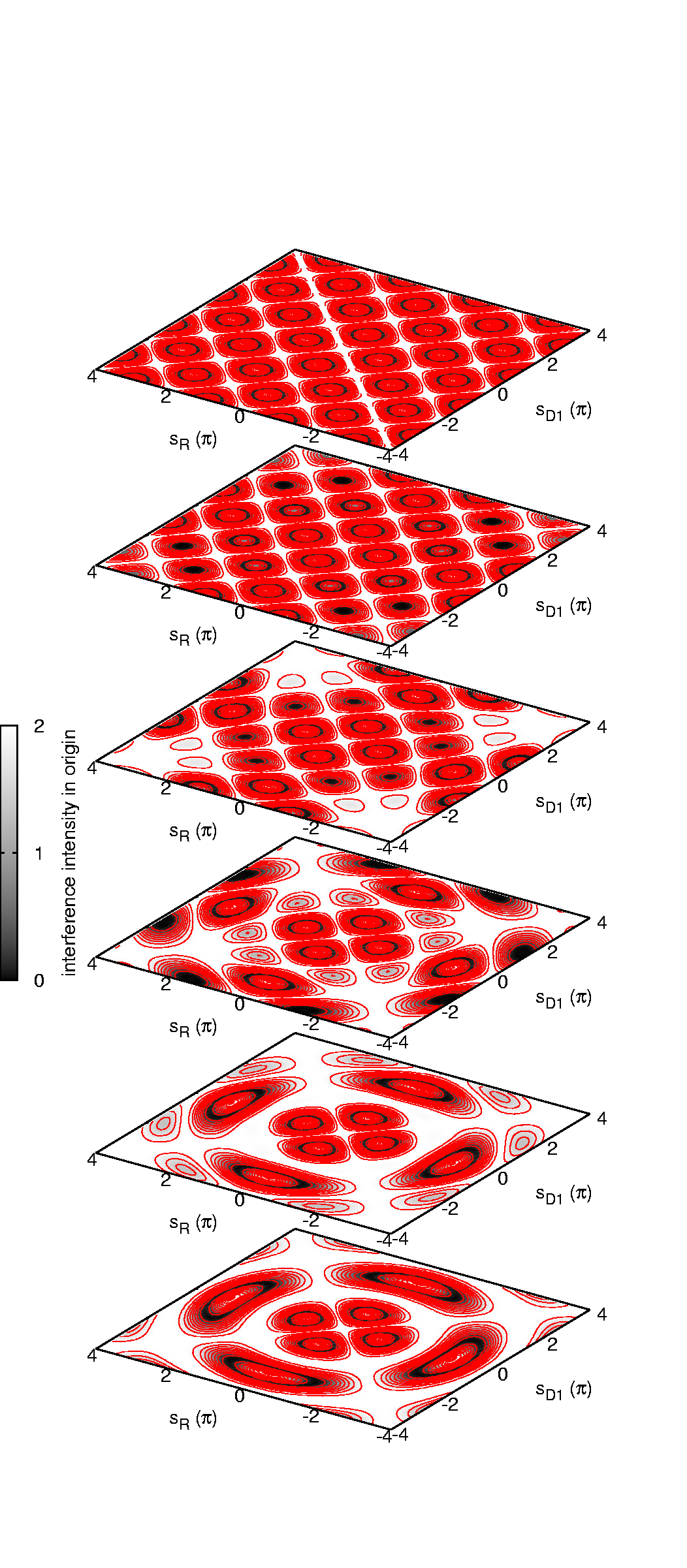}}
\caption{Interference intensity in the origin for the square as a function of the Rashba precession angle $s_R$ and the linear Dresselhaus precession angle $s_{D1}$ in units of $\pi$. From bottom to top, the square makes an angle $0,\pi/20,\pi/10,3\pi/20,\pi/5,\pi/4$ with the $\hat{\bm{x}}$-axis. The interval between contour lines is $0.25$.}
\label{fig:rtint}
\end{center}
\end{figure}

For general $\alpha_R$, $\alpha_{D1}$ the above analysis does not lead 
to a closed expression for the total rotation. The general case can 
however be easily calculated numerically by multiplying rotation 
matrices 
and plugging the result into Eq.\ (\ref{eq:averageprobability}). The result obtained for the interference intensity in the origin for the square is shown in Fig.\ \ref{fig:rtint}.
Fig.\ \ref{fig:rtint} shows the interference pattern for different 
alignments of the square in the $xy$-plane from $0$ to $\pi/4$ angle 
with the $\hat{\bm{x}}$-axis. The pattern is symmetric with respect to the 
offset angle $\pi/4$ because of the symmetry between the Rashba and linear 
Dresselhaus interaction and the pattern repeats itself of course after 
offset $\pi/2$. We observe as derived in Eq.\ (\ref{eq:aRisaD1}) the 
maxima lines 
$\alpha_R=\pm\alpha_{D1}$ and the symmetry between Rashba and linear 
Dresselhaus. The maxima and minima along the lines $s_{D1}=0$, 
$s_R\neq 0$ and $s_R=0$, $s_{D1}\neq 0$ follow from 
Eqs. (\ref{eq:rootcondition1}) and (\ref{eq:rootcondition2}) with 
$n=4$. They are given by $s_{\rm{max}}=0,\pi,2\pi $ \& $s_{\rm{min}}=\arccos(1-\sqrt{2}),2\pi-\arccos(1-\sqrt{2})$.
As we discuss in the following Section, the strong dependence 
on the precession angle and alignment can be exploited to observe the spin-interference.

\section{Discussion}

\begin{figure}[h]
\begin{center}
\resizebox{150mm}{!}{\includegraphics{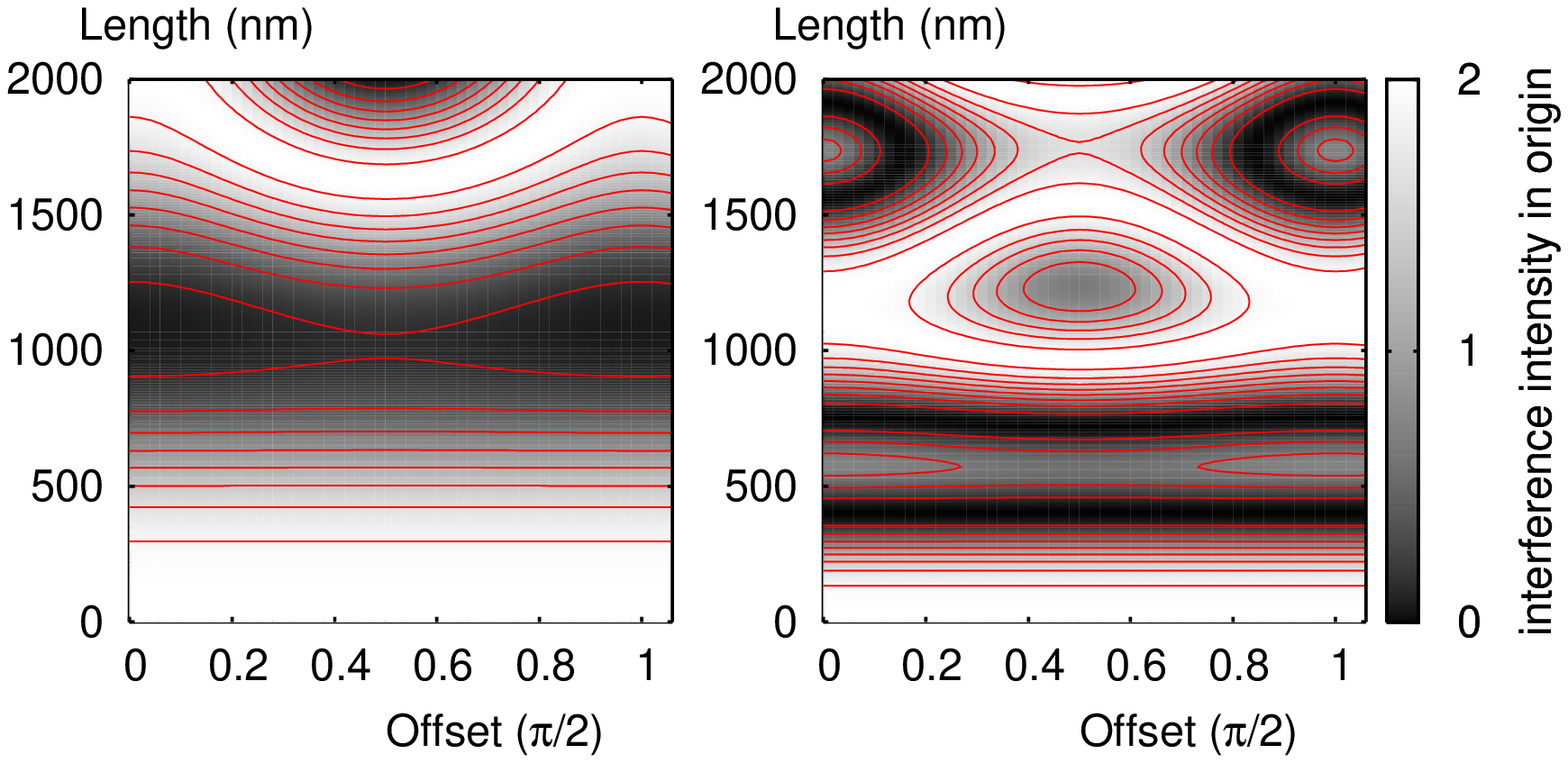}}
\caption{Interference intensity in the origin for the square as a function of the sidelength and alignment with the $\hat{\bm{x}}$-axis for linear Dresselhaus  coefficient $\alpha_{D1} = 0.7\times10^{-12}\;\rm{eVm}$ and Rashba coefficient $\alpha_R = 1.5\times10^{-12}\;\rm{eVm}$ for the left figure and $\alpha_R = 3.0\times10^{-12}\;\rm{eVm}$ for the right figure. $\rm{L} = 1000\;\rm{nm}$ corresponds to $s_{D1} = 0.39\pi$, $s_R = 0.84\pi$ respectively $s_R = 1.68\pi$ for an effective mass $m^* = 0.067\;m_0$ with $m_0$ the bare electron mass. The interval between contour lines is $0.25$.}
\label{fig:artsqmeassRsD1}
\end{center}
\end{figure}

\begin{figure}[h]
\begin{center}
\resizebox{150mm}{!}{\includegraphics{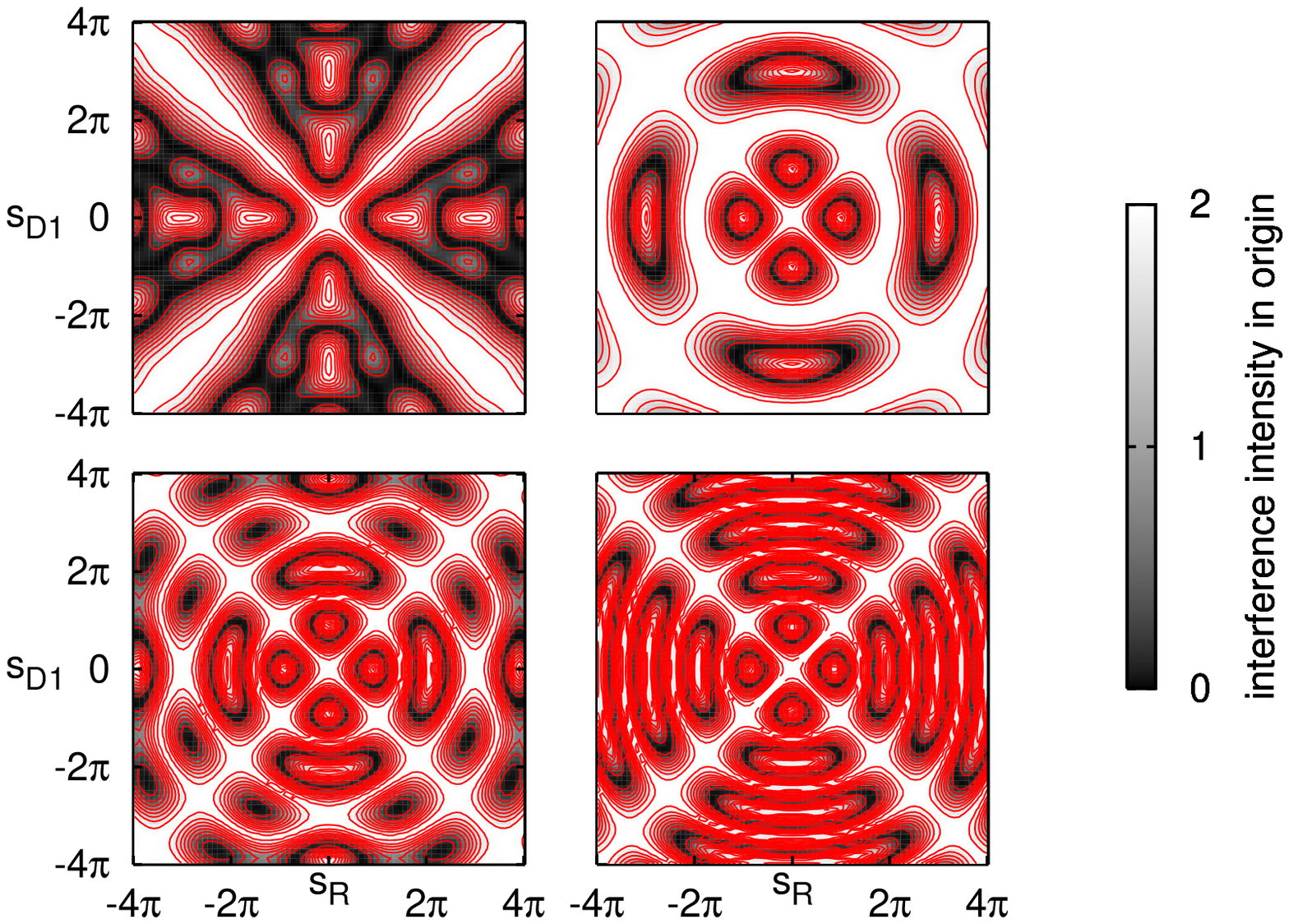}}
\caption{Interference intensity in the origin for the triangle (upper left), square (upper right), hexagon (lower left) all aligned along the $\hat{\bm{x}}$-axis and the circle (lower right). The sidelengths of the triangle and the hexagon and the radius of the circle have been scaled according to: $3L_{\rm{triangle}}= 6L_{\rm{hexagon}}= 2\pi R = 4L_{\rm{square}}$ and the units of $s_R$ and $s_{D1}$ correspond to the rotation angle over one sidelength of the square. The interval between contour lines is $0.25$.}
\label{fig:trsqhexci}
\end{center}
\end{figure}

As a first order approximation the conductance is inversely proportional to the interference intensity in the origin \cite{Buttiker85}, hence any alteration of the spin-interference should manifest itself as a change in the conductance. Furthermore, as observed from Eq.\ (\ref{eq:averageprobability}), the spin-interference affects the amplitude of the AB oscillation which is an experimentally measurable quantity in solid state devices \cite{AAS1981}. In the following discussion we will therefore assume that the interference intensity can measured and we will focus on the spin-interference itself and we will stipulate how to obtain the spin-orbit parameters from it.\newline
From Fig.\ \ref{fig:rtint} it is observed that the spin-interference varies as a function of the precession angle and the alignment of the polygon (a square in this case). The precession angle is a function of the sidelength of the polygon as well as a function of the Rashba and Dresselhaus coefficients $\alpha_i,\;i=R,D1$, see Eq. (\ref{eq:angleRashbaLength}). With both the Rashba and Dresselhaus coefficients experimentally accessible parameters and the sidelength and alignment at choice, the spin-interference can be experimentally changed. In turn, the known dependence of the spin-interference on the sidelength and the alignment can be used to find the Rashba and Dresselhaus coefficients. As an illustration we have plotted in Fig.\ \ref{fig:artsqmeassRsD1} the spin-interference of the square as a function of the sidelength and the alignment for realistic values for the Rashba and Dresselhaus constants. 
Varying the sidelength corresponds to moving along a line through zero with slope determined by the ratio of $\alpha_R$ and $\alpha_{D1}$ in one of the figures of Fig.\ \ref{fig:rtint}. From measurements with different sidelength $L$ as in Fig.\ \ref{fig:artsqmeassRsD1} it is possible to draw contour lines of the measured interference intensity through which the lines should pass. Multiple measurements will single out a specific line hence giving the ratio of the coefficients. The order of magnitude of the coefficients can be quickly determined from the number of extrema within a certain range of the sidelength $L$. To find the coefficients individually, it will be necessary to fit the interference pattern with our model. In that respect it is promising that for different values of the coefficients the spin-interference displays very distinct behavior, which is illustrated in Fig.\ \ref{fig:artsqmeassRsD1} which plots the interference pattern for two values of $\alpha_R$. \newline
Until now we have constricted the discussion to square patterns but the type of polygon can be exploited as well for the determination of the spin-orbit parameters. In Fig.\ \ref{fig:trsqhexci} we have plotted the interference pattern for the triangle (upper left), square (upper right), and hexagon (lower left), all aligned along the $\hat{x}$-axis.
The spin-orbit parameters can then be determined by measuring the interference intensity as a function of polygon which, by comparison with our model, will yield the parameters. Especially the interference patterns for polygons with low number of vertices are distinctive. With increasing number of vertices, the polygonal interference pattern converges rapidly to that of a circle shown in the lower right of Fig.\ \ref{fig:trsqhexci} (which is actually a $1000$ sided polygon), see below. \newline
In a real sample there might be two or more current leads attached to the polygon giving interference effects at all of them. However, the interference at the lead(s) different from the one through which the electron enters will be smeared out by placing a large number of polygons in a row since the interference of the space-part of the wavefunction at the lead(s) is uncorrelated between polygons. This has been nicely shown in \cite{Umbach86} to be the case for the AB $h/e$ oscillations.\newline
In general, another spin-orbit contribution to the Hamiltonian, the cubic Dresselhaus term given by
\begin{equation}
H_{D3} = \alpha_{D3}(\hat{\bm{x}}k_xk_y^2-\hat{\bm{y}}k_x^2k_y)
\end{equation}
can become important for certain heterostructures. This term can be readily included numerically. We observe however that the cubic Dresselhaus term vanishes identically for the square aligned along the $\hat{\bm{x}}$--direction and its effect can accordingly be minimized. Furthermore, from simulations it is observed that, for reasonable values of $\alpha_{D3}$, its effect on the spin-interference for the square aligned along the $\hat{\bm{x}}-\hat{\bm{y}}$-direction is minimal. Notice from Fig.\ \ref{fig:rtint} that for these two alignments of the square the spin-interference displays distinct features hence the alignment would still count as a variable.\newline
The question that arises is how robust the spin-interference effect is to deviations from the exact polygon-path. If we imagine a finite channel in the shape of a polygon in which the electron will undergo specular reflection on the walls, then in our model the path will be a zig-zag pattern and we might suspect that contributions to the rotation of the zig and zag paths will cancel each other as long as they are small. More specifically, if we choose the $\hat{\bm{y}}$-axis as the direction of the rotation corresponding to motion along the polygon side  and the direction of the zig and zag rotation under an angle $\pm\beta$ respectively with the $\hat{\bm{y}}$-axis in the $\hat{\bm{x}}-\hat{\bm{y}}$-plane, then a combined zig-zag gives a rotation
\begin{equation}
\begin{aligned}
R_{\xi'}R_{\xi} &= R_z(\beta)R_y(s')R_z(-2\beta)R_y(s')R_z(\beta)\\
&= R_y(2s')-4ie_2^2(s')e_3(\beta)\left(\begin{array}{rr}1 & 0\\ 0 &-1\end{array}\right) \\
&\quad + 4e_2(s')e_3^2(\beta)\left(\begin{array}{rr}0 & -1\\ 1 &0\end{array}\right) + O(e_2^2(s')e_3^2(\beta)),
\end{aligned}
\end{equation}
where $s' = \sqrt{s^2+w^2}$ with $s$ the precession angle along the exact polygon side and with $w$ the precession angle due to the finite width. If both the deviation angle ($e_3(\beta)$) and the spin-orbit coupling ($e_2(s')$) are small then the correction due to the finite width is third-order. This applies also to a polygon with a large number of vertices and the deviation from a circular trajectory will therefore be third order, explaining the rapid convergence of the interference pattern to that of a circle, as discussed above.\newline
If the spin-orbit coupling is finite then the finite width gives a linear correction to the rotation. Also, if the spin-orbit interaction is small and the deviation angle large, a linear correction results. Hence the calculated spin-interference effect is in general very sensitive to deviations from the exact polygon path and a finite width would result in an average over many possible paths that could potentially smear out the spin-interference.
\newline
\indent
In summary, spin-orbit coupling in polygon-structures gives rise to spin-interference with strong dependence on the spin-orbit coupling constants and sidelength and alignment of the polygon. The dependence on the sidelength, alignment, and polygon-type can be exploited to extract the spin-orbit parameters from the spin-interference pattern which is observable for instance from its effect on the amplitude of the Aharonov-Bohm oscillation. 
\newline
The first of us would like to thank Harmen Warringa for reading the article.
\bibliography{biblio}
\end{document}